# MODE-COUPLING APPROXIMATIONS, GLASS THEORY AND DISORDERED SYSTEMS.

Jean-Philippe Bouchaud, Leticia Cugliandolo

*Service de Physique de l'Etat Condensé, CEA-Saclay, Orme des Merisiers, 91 191 Gif s/ Yvette CEDEX, France*
*bouchaud, leticia@amoco.saclay.cea.fr*

Jorge Kurchan and Marc Mézard

*Laboratoire de Physique Théorique de l' Ecole Normale Supérieure* * *, 24 rue Lhomond, 75231 Paris Cedex 05, France*
*kurchan, mezard@physique.ens.fr*

## Abstract

We discuss the general link between mode-coupling like equations (which serve as the basis of some recent theories of supercooled liquids) and the dynamical equations governing mean-field spin-glass models, or the dynamics of a particle in a random potential. The physical consequences of this interrelation are underlined. It suggests to extend the mode-coupling approximation to temperatures well below the freezing temperature, in which aging effects become important. In this regime we suggest some new experiments in order to test a non-trivial prediction of the Mode-Coupling picture, which is a generalized relation between the short ($\beta$) and long ($\alpha$) time regimes.

Typeset using REVTEX

# I. INTRODUCTION

Let us face it: there are not so many techniques to deal with the score of strongly non-linear problems that Nature perversly offers, to the theoretical physicist dismay. Among others, one may of course cite fully developed turbulence[1], but also interface growth and disordered systems[2] and strongly interacting liquids (i.e. glasses)[3]. The core of most of these problems is a non-linear dynamical equation, which we write in a symbolic way as:

$$\frac{\partial \boldsymbol{\phi}(\boldsymbol{x},t)}{\partial t} = -\mu(t)\boldsymbol{\phi}(\boldsymbol{x},t) - g\boldsymbol{F}(\boldsymbol{\phi}) + \boldsymbol{\eta} \tag{I.1}$$

where $\boldsymbol{\phi}(\boldsymbol{x},t)$ is a vector field, $\boldsymbol{F}(\boldsymbol{\phi})$ is a non-linear (though generally local it can also be non-local) coupling term and $\boldsymbol{\eta}$ a Gaussian white noise. The term containing $\mu(t)$ is a restoring force. We leave open the possibility that it can become time dependent in order to include in our study the cases where one imposes a 'spherical' constraint on the field $\boldsymbol{\phi}$, such as $\boldsymbol{\phi}(t) \cdot \boldsymbol{\phi}(t) \equiv 1$. The coupling constant $g$ serves as a book-keeping parameter to set up a perturbative expansion. This expansion can either be well-behaved or ill-behaved depending – say – on the dimension of space. It is in any case rather useless when g is of order 1 if it cannot be resummed in one way or another. A very popular and versatile class of resummation schemes amounts to performing a 'one-loop' self-consistent perturbation theory. Depending on the context, self-consistent approximations of this type have received the names of 'Mode Coupling Approximation' (MCA)[4] for critical dynamics or liquids, or 'Direct Interaction Approximation' (DIA)[1] for turbulent flows; to some extent the Hartree approximation also falls in this category, as well as the refined version called 'Self Consistent Screening Approximation' (SCSA)[7]. In the MCA-DIA for the problem described by the Langevin process (I.1) one expands the relevant physical quantities to lowest non-trivial order in $g$ and then replaces the bare objects in the correction term by the fully 'renormalised' objects that one wishes to compute. This amounts to resumming a particular (infinite) set of terms in the perturbation expansion. In this way, non-trivial self-consistent equations are obtained, which enable one to peep into the strong coupling regime.

The problem is of course to try to control this procedure. An important step in this direction is to identify a model for which the self-consistent equations are *exact* (just as the Hartree approximation describes exactly the large $N$ limit of an $N$ component field problem). This is interesting for three reasons: first of all, it shows that *if* the underlying model is well behaved, the approximation does not violate any physical constraint. Second, the ingredients needed to build the model shed light into the physical content of the approximation. Third, one may hope to find a systematic expansion around this approximation. One can discuss in particular whether the interesting features of the self-consistent equations are or not an artifact of the approximation itself.

This general concern is particularly relevant within the context of supercooled liquids, for which the Mode-Coupling Theory (MCT) [†] offers (at present) the most comprehensive and

---

[†]The Mode-Coupling Theory of glasses takes as a starting point an exact Liouvillian description of

successful description[3,5]. It was understood long ago by Kraichnan[6] that the DIA approximation for turbulence becomes exact when one considers a generalisation of the Navier-Stokes equation which contains some *quenched disorder*. Recently, it has been understood that this same approximation also becomes exact for a system with deterministic, but highly chaotic interactions[8], which in fact are not very different from random ones (we shall return to this paper later on). This also holds for the simplest mode coupling equations with cubic interactions[9]. The existence of an underlying disordered problem is in fact a very general result: we shall show below that the MCA for a general non-linear $\boldsymbol{F}(\boldsymbol{\phi})$ and the dynamical generalisation of the SCSA[7] are the exact equations describing a suitably chosen disordered system.

One extra difficulty of modelling 'true' glasses (with respect to spin-glasses) is that the effective disordered potential slowing down the particles is 'self-induced' by the dynamics itself, rather than arising from an external source of quenched randomness. At the same time, glasses and spin-glasses behave very much in the same way, suggesting that the difference between 'self-induced' and quenched disorder might not be crucial, at least in a restricted time window. This scenario has been substantiated within several mean-field like models in the recent years[10–12]. In a sense, the MCT introduces some quenched randomness into the glass problem, without specifying it explicitly. This might be a clue to understand the success of the MCT.

The fact that MCT equations become exact for some disordered system suggests how to extend it to low temperatures, i.e. inside the glass phase. The MCT for glasses usually addresses the temperature regime *above* the glass transition, in the supercooled liquid phase, where the property of time translation invariance (TTI) holds. This means that the correlations between time $t$ and $t'$ depend only on the time difference $t - t'$ (as a matter of fact, the MCT is generally formulated directly in frequency space).

However, as is now well documented experimentally in the case of spin-glasses[13] and other structural glasses[14], this property *does not hold* in general in the glass phase. There is a non-vanishing 'waiting time' dependence in the correlation and response function – the 'aging' effect. It was recently observed[16] that even such simple disordered models (as the ones for which MCA is exact) have a low-temperature out of equilibrium dynamics that is both soluble and indeed captures aging phenomena in qualitative agreement with the experiments. Hence it is important to know in general how Mode-Coupling-like equations can be written in a two-time formalism, without assuming TTI, since this allows one to make predictions *deep into* the glass phase, and not just above it.

Finally, one could hope that some sort of perturbative expansion, taking the disordered system as a starting point, would bring one back closer to the original model, in particular accounting for finite dimensionality effects.

The aim of this paper is threefold. We first show that the MCA for a general $\boldsymbol{F}(\boldsymbol{\phi})$ is

---

the interacting particles but not Langevin noise $\eta$. Through a series of approximations, similar in spirit to, but different from, the MCA, one obtains the so-called Mode-Coupling equations discussed in section 4, which happen to be identical to the MCA equations deriving from Eq. (I.1). There is thus a slight distinction between MCA and MCT.

equivalent to studying a general spin-glass system (Section II). Second, we show (Section III) that Bray's SCSA for the usual $\phi^4$ theory amounts to studying a disordered version of the Bernasconi model[21], which was studied recently precisely to give some flesh to the idea of 'self-induced' disorder in glasses. Finally, we summarize in Section IV the known results[16–18] on these disordered models and rephrase them in the context of supercooled liquids. We suggest that well controlled aging experiments deep below the dynamical glass transition temperature might serve as a crucial test for the Mode-Coupling description of glasses. The reader who is interested in the physical aspects of the discussion and less by the technical details can jump directly to Section IV.

## II. MODE COUPLING APPROXIMATION AND DISORDERED SYSTEMS

We first describe the MCA on the simple case of a single scalar degree of freedom $\phi$, with an energy

$$H = \frac{\mu(t)}{2} \phi^2 + \frac{g}{4!} \phi^4 .  \qquad (II.1)$$

We assume that the dynamics of $\phi$ in contact with a heat bath is described by the Langevin equation:

$$\frac{\partial \phi}{\partial t} = -\mu(t)\phi - \frac{g}{3!} \phi^3 + \eta \qquad (II.2)$$

with initial condition $\phi(t = 0) = 0$. The thermal noise $\eta$ is a Gaussian noise $\eta$ with $\langle \eta(t) \rangle = 0$ and $\langle \eta(t)\eta(t') \rangle = 2T \delta(t-t')$ (in the following the brackets will always denote an average over the realisations of the Gaussian white noise $\eta$).

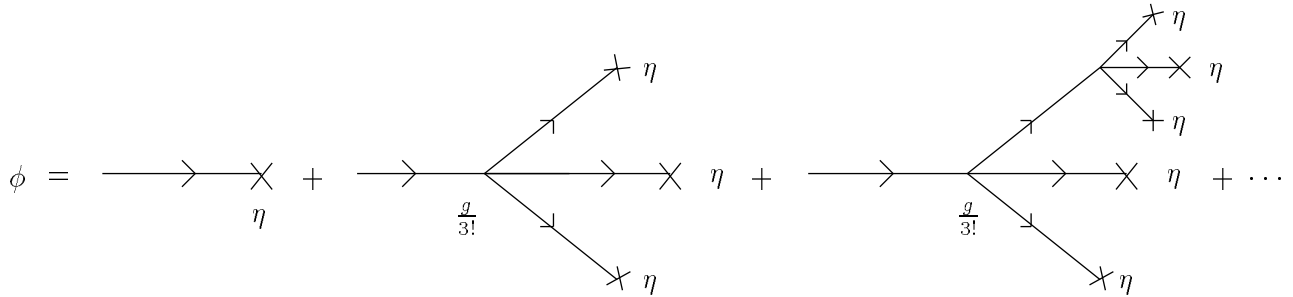

**Figure 1**. Diagrammatic representation of the perturbative solution to Eq.(II.1). Crosses indicate noise and oriented lines indicate the bare propagator $G_0$

Setting $G_0 = [\mu(t) + \frac{\partial}{\partial t}]^{-1}$, the perturbative expansion for $\phi(t)$ is easily written as:

$$\phi(t) = G_0 \otimes \eta - \frac{g}{3!} G_0 \otimes \{G_0 \otimes \eta \bullet G_0 \otimes \eta \bullet G_0 \otimes \eta\} + ... \tag{II.3}$$

where $\otimes$ means a time convolution: $(G_0 \otimes f)(t) = \int_0^t dt' G_0(t,t') f(t')$ and $\bullet$ is a simple product. For the specific form of $G_0$ in Eq. (II.2), one has $G_0(t,t') = \exp\left(-\int_{t'}^t d\tau\, \mu(\tau)\right)$. Eq. (3) can be graphically represented as in Fig 1. Crosses indicate noise and oriented lines indicate the bare propagator $G_0$.

Two quantites of interest are the (two-times) correlation function $C(t,t')$ and the response function $G(t,t')$ defined as

$$C(t,t') \equiv \langle \phi(t)\,\phi(t') \rangle, \tag{II.4}$$

$$G(t,t') \equiv \langle \frac{\partial \phi(t)}{\partial \eta(t')} \rangle = \frac{1}{2T} \langle \phi(t)\,\eta(t') \rangle. \tag{II.5}$$

where the last equality holds for a gaussian noise. The diagrammatic expansion of $C, G$ is represented in Fig. 2.

**Figure 2 - a**. Diagrammatic representation of the perturbative expansion of the auto-correlation function $C(t,t') \equiv \langle \phi(t)\phi(t') \rangle$

**Figure 2 - b**. Diagrammatic representation of the perturbative expansion of the response function $G(t,t') \equiv 1/(2T)\langle \phi(t)\eta(t') \rangle$

In what follows we shall assume that the mass is renormalised in such a way that all tadpoles (i.e. the second diagrams in Figs 2-a and 2-b) are already resummed.

It is useful to introduce the kernels $\Sigma(t,t')$ and $D(t,t')$ through the Dyson equations:

$$G(t,t') \equiv G_0(t,t') + \int_{t'}^{t} dt_1 \int_{t'}^{t_1} dt_2 \, G_0(t,t_1) \, \Sigma(t_1,t_2) \, G(t_2,t') \,, \tag{II.6}$$

$$C(t,t') \equiv \int_0^t dt_1 \int_0^{t'} dt_2 \, G(t,t_1) \, D(t_1,t_2) \, G(t',t_2) \,. \tag{II.7}$$

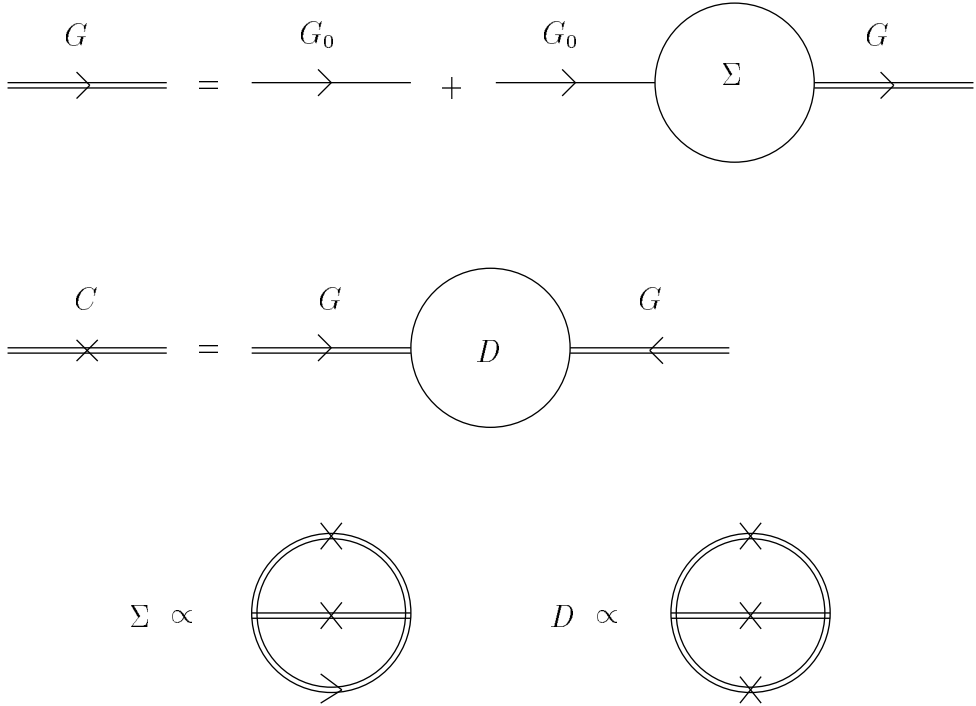

**Figure 3**. Diagrammatic representation of the MCA. The first two lines represent the Dyson equations, Eqs(II.6)-(II.7), which define the kernels $\Sigma$ and $D$. The last identity gives the value of these kernels within the MCA in the case of the $\phi^4$ theory. An oriented double line denotes the full response $G$, an oriented single line denotes the bare response $G_0$, and a crossed double line denotes the full correlation $C$.

The MCA for this problem amounts to an approximation of the kernels $\Sigma(t,t')$ and $D(t,t')$ where one takes their values at order $g^2$ and substitutes in them the bare propagator $G_0$ and the bare correlation by their renormalised values. This gives the following self-consistent equations:

$$\Sigma(t,t') = \frac{g^2}{2} C^2(t,t') G(t,t')$$

$$D(t,t') = 2T\, \delta(t-t') + \frac{g^2}{6} [C(t,t')]^3 \; , \tag{II.8}$$

which are represented in Fig. 3. This approximation neglects 'vertex renormalisation': It keeps for instance the diagram depicted in Fig. 4-a that represents a line correction, while leaving aside the diagram drawn in Fig. 4-b that represents a vertex correction.

It will also be useful in the following to note that the Dyson equations can be recast, after multiplying by $G_0^{-1}$, into the form:

$$G_0^{-1} \otimes G = \mathcal{I} + \Sigma \otimes G \; , \tag{II.9}$$

$$G_0^{-1} \otimes C = D \otimes G + \Sigma \otimes C \; , \tag{II.10}$$

where $\mathcal{I}$ is the identity operator. More explicitly,

$$\frac{\partial G(t,t')}{\partial t} = -\mu(t)\, G(t,t') + \delta(t-t') + \int_{t'}^{t} dt''\, \Sigma(t,t'')\, G(t'',t') \tag{II.11}$$

$$\frac{\partial C(t,t')}{\partial t} = -\mu(t)\, C(t,t') + \int_{0}^{t'} dt''\, D(t,t'')\, G(t',t'') + \int_{0}^{t} dt''\, \Sigma(t,t'')\, C(t'',t') \; . \tag{II.12}$$

The delta-function imposes $G(t,t^-) = 1$.

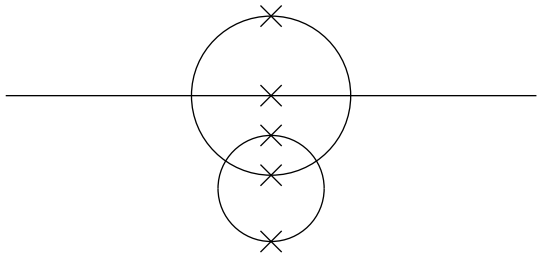

**Figure 4 - a.** Exemple of a graph that is kept in the MCA

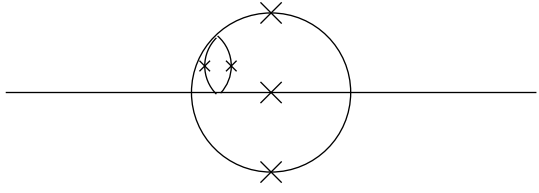

**Figure 4 - b.** Exemple of a graph that is neglected in the MCA

The basic remark is that the diagrams retained by the MCA are precisely those which survive if one considers the following disordered problem. First, one upgrades $\phi$ to an $N-$ 'colour' object $\phi_\alpha$, where $\alpha = \{1, 2, ..., N\}$. The equation of motion Eq. (II.2) is generalized to:

$$\frac{\partial \phi_\alpha}{\partial t} = -\mu(t)\, \phi_\alpha - 4g \sum_{\beta < \gamma < \delta} J_{\alpha\beta\gamma\delta}\ \phi_b \phi_\gamma \phi_\delta + \eta_\alpha \qquad (\text{II}.13)$$

with independent noises $\eta_\alpha$. This equation derives from the Hamiltonian

$$H_J = g \sum_{\alpha < \beta < \gamma < \delta} J_{\alpha\beta\gamma\delta}\ \phi_\alpha \phi_\beta \phi_\gamma \phi_\delta \qquad (\text{II}.14)$$

The couplings $J_{\alpha\beta\gamma\delta}$ are independent Gaussian random variables of zero mean and variance $\overline{J^2_{\alpha\beta\gamma\delta}} = 1/N^3$). In the large $N$ limit, the correlation:

$$C(t,t') \equiv \frac{1}{N} \sum_{\alpha=1}^{N} \overline{\langle \phi_\alpha(t) \phi_\alpha(t') \rangle} \qquad (\text{II}.15)$$

(where the overline denotes the average over the random couplings $J_{\alpha\beta\gamma\delta}$) and the response:

$$G(t,t') \equiv \frac{1}{N} \sum_{\alpha=1}^{N} \overline{\langle \frac{\partial \phi_\alpha(t)}{\partial \eta_\alpha(t')} \rangle} \qquad (\text{II}.16)$$

precisely obey the MCA equations, Eqs. (II.7,II.8,II.6). (It is important to notice that the random couplings are *quenched*, i.e. time-independent random variables.) The fact that MCA equations are recovered can be seen either directly on the perturbation theory, or through the use of functional methods given in Appendix A. A simple physical interpretation can be obtained through the cavity method[19,17] where one shows that, in the large $N$ limit, any one of the $\phi_\alpha$ evolves through an effective *linear* Langevin equation:

$$\frac{\partial \phi_\alpha}{\partial t} = -\mu(t)\phi_\alpha + \int_0^t \Sigma(t,t')\phi_\alpha(t') + \xi_\alpha(t) + \eta_\alpha(t) \tag{II.17}$$

where $\xi_a(t)$ is an effective (Gaussian) noise, with correlations self-consistently given by $\langle \xi_a(t)\xi_a(t')\rangle = D(t,t')$. This result, derived in detail in Appendix A, gives back the MCA equations for the two point functions. It also provides a precise recipe to calculate higher order correlation functions within the MCA.

The first to notice that the MCA (for the case of a 'quadratic' dynamical equation) corresponds to the exact dynamical equations of a disordered problem with a large number of components was Kraichnan[6] in the context of the Navier-Stokes equation. The important property of the random couplings which is used in the derivation is that the couplings are independent Gaussian variables.

In the case of $J$'s with three indices, this can also be implemented using a deterministic construction of the $J_{\alpha\beta\gamma}$, in terms of Clebsch-Gordan coefficients of an $O(3)$ symmetry group [‡]. This was first noticed by Amit and Roginsky[22], and has been recently extended for dynamical problems,[8,23,9].

Interestingly enough, this equivalence between MCA and a disordered system extends to an arbitrary non-linear coupling $F(\phi)$ (see Eq. (1)). Expanding $F(\phi)$ in power series

$$F(\phi) = \sum_{r=2}^{\infty} \frac{F_r}{r!}\phi^r \tag{II.18}$$

the natural generalisation of the MCA (i.e. neglecting all vertex renormalisation) reads:

$$\Sigma(t,t') = g^2 \sum_{r=2}^{\infty} \frac{F_r^2}{(r-1)!}[C(t,t')]^{r-1} G(t,t'), \tag{II.19}$$

$$D(t,t') = 2T\delta(t-t') + g^2 \sum_{r=2}^{\infty} \frac{F_r^2}{r!}[C(t,t')]^r. \tag{II.20}$$

(Note that for $r$ odd, there appears an additional 'tadpole' contribution in Eq. (II.19), which we have assumed again, that it has been reabsorbed into the mass term $\mu(t)$). The dynamical equations within the MCA for this extended model are readily obtained inserting these expressions for $\Sigma$ and $D$ in Eqs. (II.11) and (II.12).

These equations can again be obtained as the exact solution of a problem with quenched randomness, the problem of $N$ continuous spins $\phi_\alpha$ interacting through the Hamiltonian:

---

[‡]The behaviour of these coefficients as a function of their indices looks however extremely 'chaotic' – the difference between determinism and randomness is thus probably very thin. See the discussion in Ref. [11].

$$H_J[\phi] = g \sum_{r \geq 2}^{\infty} F_r \sum_{\alpha_1 < ... < \alpha_{r+1}} J_{\alpha_1 ... \alpha_{r+1}} \phi_{\alpha_1} ... \phi_{\alpha_{r+1}} \qquad (\text{II.21})$$

and the Langevin equation

$$\frac{\partial \phi_\alpha}{\partial t} = -\mu(t) \phi_\alpha - \frac{\delta H_J[\phi]}{\delta \phi_\alpha} + \eta_\alpha \qquad (\text{II.22})$$

where $J_{\alpha_1,..\alpha_{r+1}}$ are quenched symmetrical and otherwise independent Gaussian variables normalized as:

$$\overline{(J_{\alpha_1,..\alpha_{r+1}})^2} = \frac{1}{N^r}. \qquad (\text{II.23})$$

Therefore the mode coupling equations corresponding to an arbitrary nonlinearity $F(\phi)$ describe exactly a spin-glass problem with arbitrary multispin interactions. Let us note that in order to be well defined, the model defined by the Hamiltonian $H_J$ must be supplemented by a constraint preventing the field $\phi_\alpha$ from exploding in an unstable direction set by the coupling tensor $J_{\alpha_1...\alpha_{r+1}}$. A convenient constraint is

$$\frac{1}{N} \sum_{\alpha=1}^{N} \phi_\alpha^2(t) = C(t,t) \equiv 1 \qquad (\text{II.24})$$

which can be implemented dynamically through a Lagrange multiplier, acting as a time-dependent mass $\mu(t)$ which must be self-consistently determined. Another possible regularisation is to add to $H_J$ a term $Ng'/2(\sum_\alpha \frac{\phi_\alpha^2}{N})^{\frac{r+1}{2}}$, with $g'$ large enough. As a matter of fact, this term precisely generates, for $r$ odd, the tadpole contribution in the expansion of the original $\phi^{r+1}$ model, provided one chooses $g' = g/(2^{\frac{r-1}{2}}(\frac{r+1}{2})!)$. Surprisingly, it can be checked that this value of $g'$ is not large enough to suppress the instability of the disordered model. We are thus led to conclude that the plain MCA approximation (i.e. without imposing an extra constraint) for – say – the $\phi^4$ model leads to spurious instabilities, at least at low temperatures. A similar conclusion was reached in Ref. [26].

Interestingly enough, the disordered multispin Hamiltonian can also be seen[29,17,18] as describing a particle evolving in an $N$ dimensional space in a quenched random potential $H_J[\phi] = V[\phi]$ This random potential has a Gaussian distribution with zero mean and variance[§]:

$$\overline{V(\phi)V(\phi')} = Ng^2 \sum_{r \geq 2}^{\infty} \frac{F_r^2}{(r+1)!} \left(\frac{\phi \cdot \phi'}{N}\right)^{r+1} = N\hat{\mathcal{V}}\left(\frac{\phi \cdot \phi'}{N}\right) \qquad (\text{II.25})$$

with

$$\hat{\mathcal{V}}(x) = g^2 \sum_{r=2}^{\infty} \frac{F_r^2}{(r+1)!} x^{r+1} . \qquad (\text{II.26})$$

---

[§]Note that the sign of $\mathcal{V}$ differs from the convention adopted in Ref.[18]

The general mode coupling equations (II.11),(II.12),(II.19),(II.20), are thus also the exact dynamical equations for the problem of a particle in a random potential in large dimension $N$. In this last context, they are often written[17,18] in a differential form obtained after applying the operator $G_0^{-1}$:

$$\frac{\partial C(t,t')}{\partial t} = -\mu(t)\, C(t,t') + 2T\, G(t',t) +$$
$$+ \int_0^{t'} dt''\, \hat{\mathcal{V}}'[C(t,t'')]\, G(t',t'') + \int_0^{t} dt''\, G(t,t'')\, \hat{\mathcal{V}}''[C(t,t'')]\, C(t'',t') \quad (\text{II}.27)$$

$$\frac{\partial G(t,t')}{\partial t} = -\mu(t)\, G(t,t') + \delta(t-t') + \int_{t'}^{t} dt''\, G(t,t'')\, \hat{\mathcal{V}}''[C(t,t'')]\, G(t'',t') \quad (\text{II}.28)$$

The physical consequences of this general equivalence will be fully discussed in Section IV. The extension of the mapping to a space dependent $\phi(\boldsymbol{x},t)$ (or to a multicomponent field) is straightforward. Several interesting physical examples involve an equation of the type:

$$\frac{\partial \hat{\phi}(\boldsymbol{k},t)}{\partial t} = -(\nu k^2 + \mu)\hat{\phi}(\boldsymbol{k},t) - \sum_{r=2}^{\infty} \sum_{\boldsymbol{k}_1,..\boldsymbol{k}_r} \frac{F_r}{r!}\mathcal{L}_r(\boldsymbol{k}|\boldsymbol{k}_1,.....\boldsymbol{k}_r)\hat{\phi}(\boldsymbol{k}_1,t)....\hat{\phi}(\boldsymbol{k}_r,t) + \eta(\boldsymbol{k},t)$$
$$(\text{II}.29)$$

where $\hat{\phi}(\boldsymbol{k},t)$ is the Fourier transform of $\phi(\boldsymbol{x},t)$, and $\eta(\boldsymbol{k},t)$ a Gaussian noise such that $\langle \eta(\boldsymbol{k},t)\eta(\boldsymbol{k}',t')\rangle = 2T(\boldsymbol{k})\delta(\boldsymbol{k}+\boldsymbol{k}')\delta(t-t')$. The case of the KPZ equation[2] corresponds to $r=2$, $\mathcal{L}_2(\boldsymbol{k}|\boldsymbol{k}_1,\boldsymbol{k}_2) = [\boldsymbol{k}_1 \cdot \boldsymbol{k}_2]\, \delta(\boldsymbol{k}_1+\boldsymbol{k}_2+\boldsymbol{k})$, while domain coarsening in the $\phi^4$ theory corresponds to $r=3$, $\mathcal{L}_3(\boldsymbol{k}|\boldsymbol{k}_1,\boldsymbol{k}_2,\boldsymbol{k}_3) = \delta(\boldsymbol{k}_1+\boldsymbol{k}_2+\boldsymbol{k}_3+\boldsymbol{k})$, with a negative $\mu$[24]. The Navier-Stokes equation is similar to the KPZ case, with however an extra tensorial structure due to the vector character of the velocity field[1].

The correlation and response functions now become $\boldsymbol{k}$ dependent:

$$\delta(\boldsymbol{k}+\boldsymbol{k}')C(\boldsymbol{k},t,t') = \langle \hat{\phi}(\boldsymbol{k},t)\hat{\phi}(\boldsymbol{k}',t')\rangle \quad (\text{II}.30)$$

$$\delta(\boldsymbol{k}+\boldsymbol{k}')G(\boldsymbol{k},t,t') = \langle \frac{\partial \hat{\phi}(\boldsymbol{k},t)}{\partial \eta(\boldsymbol{k}',t')}\rangle \quad (\text{II}.31)$$

The generalized MCA equations then read (assuming that the structure factors $\mathcal{L}_r(\boldsymbol{k}|\boldsymbol{k}_1,.....\boldsymbol{k}_r)$ are invariant under the permutation of $\boldsymbol{k}_1,...,\boldsymbol{k}_r$):

$$\Sigma(\boldsymbol{k},t,t') = g^2 \sum_{r=2}^{\infty} \frac{F_r^2}{(r-1)!} \sum_{\boldsymbol{k}_1,..\boldsymbol{k}_r} \mathcal{L}_r(\boldsymbol{k}|\boldsymbol{k}_1,.....\boldsymbol{k}_r)\mathcal{L}_r(\boldsymbol{k}_r|\boldsymbol{k}_1,.....\boldsymbol{k})$$
$$C(\boldsymbol{k}_1,t,t')...C(\boldsymbol{k}_{r-1},t,t')G(\boldsymbol{k}_r,t,t') \quad (\text{II}.32)$$

$$D(\boldsymbol{k},t,t') = 2T(\boldsymbol{k})\, \delta(t-t') + g^2 \sum_{r=2}^{\infty} \frac{F_r^2}{r!} \sum_{\boldsymbol{k}_1,..\boldsymbol{k}_r} (\mathcal{L}_r(\boldsymbol{k}|\boldsymbol{k}_1,.....\boldsymbol{k}_r))^2$$
$$C(\boldsymbol{k}_1,t,t')...C(\boldsymbol{k}_r,t,t') \quad (\text{II}.33)$$

where $\Sigma(\boldsymbol{k},t,t')$ and $D(\boldsymbol{k},t,t')$ are defined in analogy with Eqs. (II.6),(II.7).

## III. SCSA AND DISORDERED SYSTEMS

Another useful resummation scheme is the 'Self-Consistent Screening Approximation' introduced by Bray in the context of the static $\phi^4$ theory[7]. It amounts to using an $n$ component vector field $\phi$ and resumming self consistently all the diagrams appearing in the large $n$ expansion ($n$ is the number of components of $\phi_n$), including those of order $\frac{1}{n}$. This approximation can also be seen as a MCA when one introduces an auxiliary field by rewriting the Langevin equation for the $\phi^4$ theory as:

$$\frac{\partial \phi(t)}{\partial t} = -\mu \phi(t) - \hat{g}\phi(t)\sigma(t) + \eta_\phi \tag{III.1}$$

$$\sigma(t) = \frac{\phi(t)^2}{2} \tag{III.2}$$

with $\hat{g} \equiv \frac{2g}{3!}$. (The factor 2 has been introduced for later convenience). In this form one gets back a problem similar to the ones studied before, which can be seen as two coupled fields $\phi$ and $\sigma$ evolving with bare evolution operators

$$(G_{\phi 0})^{-1} = \mu(t) + \frac{\partial}{\partial t} \quad ; \quad (G_{\sigma 0}) = \mathcal{I} \tag{III.3}$$

Once this fictitious decomposition of the non-linear coupling is performed, one can apply the MCA on the coupled equations (III.2). Of course, if MCA were exact, the approximation would give the same results as in the previous paragraph. The fact that it is only approximate leaves room to a certain freedom on the starting point to improve (or deteriorate) the quality of the approximation (see Ref. [26] for a related discussion). Introducing two correlation functions $C_\phi(t,t')$ and $C_\sigma(t,t')$, and two response functions $G_\phi$, $G_\sigma$, together with the corresponding kernels $\Sigma_\phi$, $D_\phi$, $\Sigma_\sigma$ and $D_\sigma$, defined (separately for each field $\phi$ or $\sigma$) as:

$$G^{-1} = (G_0)^{-1} - \Sigma \quad ; \quad C = G \otimes D \otimes G^T \tag{III.4}$$

one finds the following result for the kernels[**]:

$$\Sigma_\phi(t,t') = \hat{g}^2 C_\sigma(t,t') G_\phi(t,t') - \hat{g} C_\phi(t,t') G_\sigma(t,t') \tag{III.5}$$

$$D_\phi(t,t') = 2T\delta(t-t') + \hat{g}^2 C_\phi(t,t') C_\sigma(t,t') \tag{III.6}$$

$$\Sigma_\sigma(t,t') = -\hat{g} C_\phi(t,t') G_\phi(t,t') \tag{III.7}$$

$$D_\sigma(t,t') = \frac{1}{2}[C_\phi(t,t')]^2 \tag{III.8}$$

It turns out that, again, these dynamical equations are exact for a certain (mean-field like) spin-glass model. Let us define the following 'spin-glass' Hamiltonian:

---

[**]We have again reabsorbed the tadpole contribution in $\mu(t)$

$$\mathcal{H} = \frac{J_0}{2N} \sum_{\lambda=1}^{N} [\sum_{\alpha<\beta}^{N} J_{\alpha,\beta}^{\lambda} \phi_\alpha \phi_\beta]^2 + \frac{\mu}{2} \sum_\alpha \phi_\alpha^2 \qquad (\text{III.9})$$

where the $J_{\alpha,\beta}^{\lambda}$ are identically distributed independent (apart from a constraint of symmetry *in the two indices* $\alpha, \beta$) random variables, such that $J_{\alpha,\beta}^{\lambda}$ equals 1 with probability $\frac{1}{N}$ and zero otherwise. This model was proposed and studied in Ref. [11], as a disordered proxy of the Bernasconi model (which serves as a model for glassy behaviour without randomness). The calculations showing that the dynamics of the model defined by Eq. (III.9), with $J_0 \equiv \hat{g}$, exactly reproduces the self-consistent dynamical equations (III.8) are given in Appendix B. The relation with Bray's SCSA can be directly seen on the statics of the disordered Hamiltonian, Eq. (III.9). It is straightforward to show that, in equilibrium,

$$C_\sigma(t,t) = \frac{T}{1 + \frac{\hat{g}}{2T}[C_\phi(t,t)]^2} \qquad (\text{III.10})$$

$$C_\phi(t,t) = \frac{T}{\mu + \frac{\hat{g}}{T} C_\sigma(t,t) C_\phi(t,t)} \qquad (\text{III.11})$$

which indeed coincide with Bray's equations in zero dimensions with the identification $n = 1$, and his choice for $\frac{\hat{g}}{T} = 2$.

The tadpole term in the expansion of the $\phi^4$ theory can also be taken care of by adding to the disordered system's Hamiltonian (III.9) a term in $g'/2N(\sum_\alpha \phi_\alpha^2)^2$. We notice that within this approximation the energy (III.9) is always positive, which ensures that the dynamical version of the SCSA are well defined, at variance with the MCA (cf. above). If the quadratic term in the original hamiltonian is positive, then the spin-glass sytem is unfrustrated: it has a single ground state at $\phi_\alpha = 0$. If instead we consider a double well $\phi^4$ theory with a negative $\mu$, we find a frustrated spin-glass system. The usual Bernasconi model involves *Ising* spins with the same coupling as in (III.9). It is recovered here in the limit where $\mu \to -\infty$. The study of the physical content of this dynamical SCSA is left for future work[25].

## IV. PHYSICAL DISCUSSION: MODE COUPLING BELOW $T_G$

We have shown in the previous sections that the Mode-Coupling Approximation (or the SCSA) for a non-linear dynamical Langevin equation amounts to studying an auxiliary Langevin process for a system with quenched disorder. In particular, the MCA for the Langevin process (I.1) described by the non linearity $F(\phi)$ leads to the pair of coupled dynamical equations for the correlation $C(t,t')$ and the response $G(t,t')$ written in (II.28). As we have seen, these equations describe exactly the dynamics of a particle in a random potential in a large dimensional space, or else as a certain type of mean-field spin-glass system with multispin couplings.

Actually the usual mode coupling equations which have been used successfully in the study of supercooled liquids are a special case of these general equations. The MCT is written in terms of the density-density correlation function which is normalized to one at equal times, i.e. $C(t,t) = 1$, corresponding to the spherical constraint (II.24). In the case of a supercooled liquid one studies a system in its high temperature phase where it obeys *time translation invariance* (TTI), together with the Fluctuation-Dissipation Theorem (FDT). The first of these properties allows to write the correlation and response as functions of time differences only: $C(t,t') = C(t-t')$ and $G(t,t') = G(t-t')$. The FDT, which states that

$$G(\tau) = -\frac{1}{T} \Theta(\tau) \partial_\tau C(\tau) , \qquad (IV.1)$$

(where $\tau \equiv t - t'$), enables one to rewrite the mode coupling equations as a single equation:

$$\partial_\tau C(\tau) = -\hat{\mu}_\infty C(\tau) + \frac{1}{T} \int_0^\tau d\tau'' \, \hat{\mathcal{V}}'[C(\tau - \tau'')] \, \partial_{\tau''} C(\tau'') , \qquad (IV.2)$$

where $\hat{\mu}_\infty = \lim_{t \to \infty} \mu(t) - 1/T \, \hat{\mathcal{V}}'(1)$.

Eq. (IV.2) is basically the general Mode-Coupling equation for the density correlations in a supercooled liquid above the dynamical transition temperature introduced by Leutheusser, Götze and others[3] as a 'schematic' model for the ideal glass transition. This similarity (in the high temperature phase) was already pointed out in[27,28]. The only difference lies in the fact that the Mode-Coupling equations also possess an 'inertial' term $\partial_\tau^2 C(\tau)$. In the notation of Ref. [3], the '$F_r$' models correspond to the case where the non linearity is a pure power law, where only $F_r$ is non zero; the $F_{r_1,r_2}$ models correspond to a nonlinearity which is a sum of two powers with $F_{r_1}, F_{r_2}$ different from zero, etc. One should note that the Mode-Coupling equations for supercooled liquids were written *from the start* within a TTI formalism [††]. The analysis of these mode coupling equations (IV.2) for supercooled liquids has shown the existence of a dynamical phase transition at a certain temperature $T_d$ (which is traditionally called $T_c$ in the MCT), and identified two classes of behaviours (called A and B) when the temperature decreases and approaches $T_d$. This same classification has also been discussed in the spin-glass dynamics framework, where the temperatures lower than $T_d$ (i.e. inside the spin-glass phase) has also been discussed and has led recently to several interesting developments. As noted recently in Ref. [9], these studies of spin-glass dynamics below $T_d$ provide a natural generalisation of the mode coupling equations below the glass temperature, the physical content of which we shall discuss.

Let us thus summarize the important results associated to the dynamical equations (II.28) and rephrase them in the context of the Mode-Coupling theory.

There exists a critical temperature $T_d$ (or a set of coupling constants $F_r$) separating a 'liquid' (or paramagnetic) phase where TTI and FDT hold. The dynamics is described by Eq. (IV.2) and the correlations decay to zero at large times: $C(\tau) \to 0$ when $\tau \to \infty$. The transition can be of two types. In a first class of systems the transition is a continuous one: the analysis of the static situation through the replica method leads to a 'continuous replica symmetry breaking'[19] transition occuring at the temperature $T_s$ which coincides with the dynamical temperature $T_d$ where the ergodicity is broken. This corresponds to class A in the classification of Ref. [3]. The second class of systems have a very different behaviour where the static transition temperature $T_s$ is smaller than the dynamical one $T_d$. This static transition, in the replica language, is a 'one step replica symmetry breaking' transition, which means that it is a first order transition from the point of view of the order parameter (but it is second order from the thermodynamic point of view). It corresponds to class B in the classification of Ref. [3]. We shall concentrate on this second category, which is supposed to be the most relevant for a description of the structural glass transition. In that respect, it is interesting to remark that class B systems correspond, in the equivalence with a particle in a random potential, to the case of *short range* correlations of random potential, whereas class A systems correspond to *long range* correlations[32,17,18].

---

[††]If one attempts to extend directly (IV.2) to the low temperature phase keeping TTI and FDT as in Ref. [3], one obtains a theory yielding *different* predictions, the meaning of which is not clear to us.

Before describing the quantitative feature of the dynamical transition for class B systems, a few comments on their physical relevance is in order. The existence of a dynamic transition above the static one is associated with the appearance of many metastable states and a breaking of ergodicity at $T_d$, which does not reflect onto the equilibrium (Gibbs) measure[28]. However, this effect can exist only at the mean-field level, and it has been suggested that in finite dimensions some nucleation processes[10,31] smooth the transition at $T_d$ and replace it by a crossover temperature range where the relaxation times will increase very fast with decreasing temperature. The glass transition temperature $T_g$, empirically defined by the fact that the relaxation time (or the viscosity) reaches a certain conventional value, would therefore lie *below* the mean-field $T_d$ (but above the static transition temperature $T_s$). Actually, the same type of argument has been developed in the study of supercooled liquids, where some 'activated processes' are supposed to smooth out the dynamical transition[3].

Hereafter we shall first recall the existing results for the dynamics above $T_d$ in spin-glasses and in supercooled liquids. These lead to the well-known predictions of the mode coupling theory for the relaxation just above $T_d$, which have been tested experimentally. Then we shall recall the results of spin-glass dynamics below the dynamical transition. These lead to some predictions for the (off equilibrium) dynamics which should apply to glasses at much lower temperatures (smaller than $T_g$), such that the relaxation time is larger than the experimental time scale.

– For $T > T_d$, the analysis of Eq. (IV.2) is sufficient. One finds that[29], for $T$ close to (but above) $T_c$, $C(\tau)$ has the form given in Fig. 5, with a plateau and the celebrated $\alpha$ and $\beta$ regimes, characterized by two exponents $\alpha$ and $\beta$ related through[3]:

$$\frac{\Gamma^2[1+\alpha]}{\Gamma[1+2\alpha]} = \frac{\Gamma^2[1-\beta]}{\Gamma[1-2\beta]} = \frac{T}{2} \frac{\hat{\mathcal{V}}'''(q)}{(\hat{\mathcal{V}}''(q))^{3/2}} \ , \qquad (IV.3)$$

with the value of the correlation at the plateau $q$ given by $(1-q)^2 \hat{\mathcal{V}}''(q) = T_c^2$. Note that these two exponents $\alpha, \beta$ are usually called $b, a$ in the Mode-Coupling litterature. We however feel that it is more appropriate to call $\alpha$ the exponent corresponding to the $\alpha$ peak, and $\beta$ the one corresponding to the $\beta$ peak!

– For $T < T_d$, there appear diverging relaxation times in the problem. It has been realised recently[16] that in this case one needs to take into account carefully the existence of an initial time for the dynamics. Stated differently, one must abandon TTI, as the age of the system becomes an important time scale in the problem. This leads to the existence of so called aging effects which have been observed in spin-glasses[13], polymer glasses[14] and also in a variety of other systems[33]. A study of the full dynamical equations (II.28) shows that one must also abandon the FDT. The system is out of equilibrium, but one can nevertheless obtain some information on its behaviour. The correlation function $C(t, t')$ (and similarly $G(t, t')$) must be decomposed into *two* parts (see Fig. 6): $C(t_w + \tau, t_w) = C_{FDT}(\tau) + \mathcal{C}(t_w + \tau, t_w)$, $C_{FDT}$ is TTI, it is related to $G_{FDT}$ through the FDT, Eq.(IV.1), and corresponds to the high frequency dynamics ($\beta$ peak), while the *aging part* $\mathcal{C}(t_w + \tau, t_w)$ is a function of the ratio $\lambda = h(t_w + \tau)/h(t_w)$ only. The 'effective time' $h$ is still not determined theoretically, but a likely possibility, advocated in[15], is that $h(t) = t$. In other words, the relaxation time corresponding to the aging part of the correlation is the experimental waiting time $t_w$ itself. The $\alpha$ regime *thus still exists for $T < T_c$ if the waiting time is finite*. Only in the limit $t_w \to \infty$ will the correlation relax to a non zero value. This is the 'weak ergodicity breaking' scenario proposed in Refs. [15],[16]: $\lim_{\tau \to \infty} \lim_{t_w \to \infty} C(\tau + t_w, t_w) = q$ and $\lim_{\tau \to \infty} C(\tau + t_w, t_w) = 0$, $\forall t_w$ finite. The exponents $\alpha$ and $\beta$ are thus still well defined for finite $t_w$ (see Fig 6) as

$$C(t_w + \tau, t_w) \sim q + c_\beta \tau^{-\beta} \qquad \text{if} \quad C \gtrsim q \qquad (IV.4)$$

$$C(t_w + \tau, t_w) \sim q - c_\alpha \left(\frac{\tau}{T_w}\right)^\alpha \qquad \text{if} \quad C \lesssim q . \qquad (IV.5)$$

Aging is manifested in the $t_w$-dependence of $T_w = dt_w/d\ln(h(t_w))$, which is an increasing function of $t_w$. In the simple case where $h(t_w) = t_w$, one has $T_w = t_w$.

The exponents $\alpha$ and $\beta$ are now given by a *modified relation* which reads[18]:

$$x \frac{\Gamma^2[1+\alpha]}{\Gamma[1+2\alpha]} = \frac{\Gamma^2[1-\beta]}{\Gamma[1-2\beta]} = \frac{T}{2} \frac{\hat{\mathcal{V}}'''(q)}{(\hat{\mathcal{V}}''(q))^{3/2}} \qquad (IV.6)$$

with $q$ given by $(1-q)^2 \hat{\mathcal{V}}''(q) = T^2$. $x$ is a temperature dependent number, $0 < x < 1$. A crucial observation is the fact that this number is not arbitrary and could be in principle measured. It actually provides the quantitative measure for the violation of FDT. More precisely, $x$ is defined as[16]:

$$\mathcal{G}(t, t') = \frac{x}{T} \frac{\partial \mathcal{C}(t, t')}{\partial t'} , \qquad (IV.7)$$

where we assume $t' < t$. The usual FDT relation would state that $x = 1$. Glassy dynamics below $T_g$ gives a value $x < 1$, which also governs the relation between the exponents $\alpha$ and $\beta$ in (IV.6).

We shall not expand here on the case of class A situations, but just mention that the behaviour in the low temperature phase is more complicated[16–18]. The correlation and response have to decomposed into *two* parts as in class B situations but the behaviour of the aging parts $\mathcal{C}, \mathcal{G}$ cannot be characterised by a single function $h(t)$ and the violation of FDT is given, in the limit of large times, by a non-trivial function of the correlation function $X[\mathcal{C}]$ (instead of the single constant $x$).

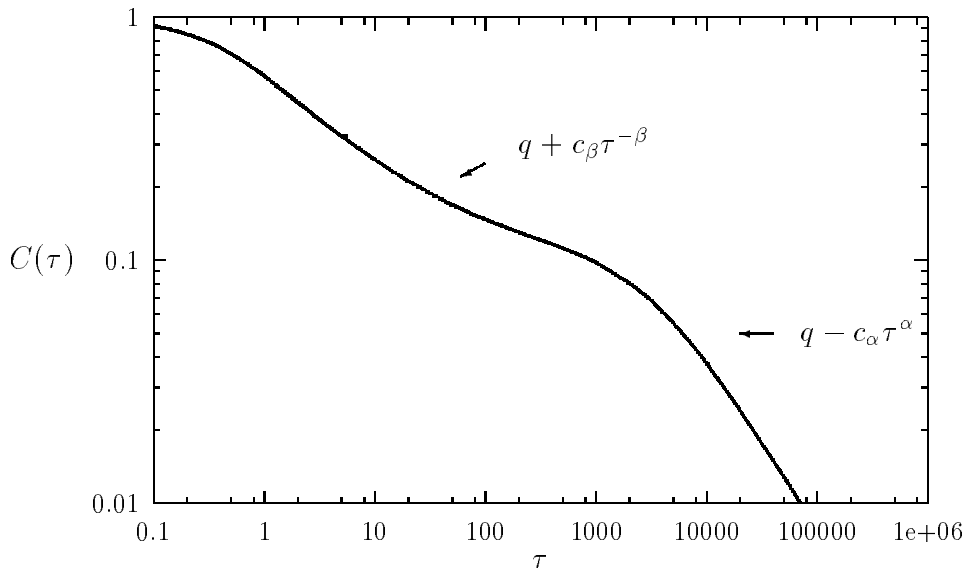

FIG. 5. Decay of the auto-correlation function above the critical temperature. $C(\tau + t_w, t_w) = C(\tau)$ vs $\tau$.

Let us finally say a word on the distinction between *explicit* and *spontaneous* non equilibrium. Throughout this paper we have discussed extensions of mode-coupling-like dynamical equations which reduce to the usual ones if one assumes that TTI and FDT hold, as when the system they describe has achieved equilibration in some component of phase-space after some finite transient. However, we now know that such equations may admit a low-temperature glassy phase in which the equilibration time is infinite: there is violation of TTI and FDT at arbitrarily long times. The reason why this *spontaneous* non-equilibrium happens is that the equilibration time diverges, or at least becomes extremely large, with the system size. On the other hand, there are systems such as surface-growth (described by the KPZ equation mentioned above) and stirred turbulence which are *by construction* non-equilibrium situations; their equations of motion do not admit any equilibrium solution even for a finite system. One can then wonder how to recognize if a given set of equations for response and correlation functions has explicit or spontaneous long-time non-equilibrium. It is interesting to notice that this question has a clear meaning within the supersymmetrical field theory described in Appendix A for the dynamics of disordered system. Any Langevin process that derives from a potential automatically yields an action that possesses a certain (super)symmetry (spontaneously broken if there is a glassy phase), while systems with *explicit* non-equilibrium have a dynamical action that break this symmetry explicitly.

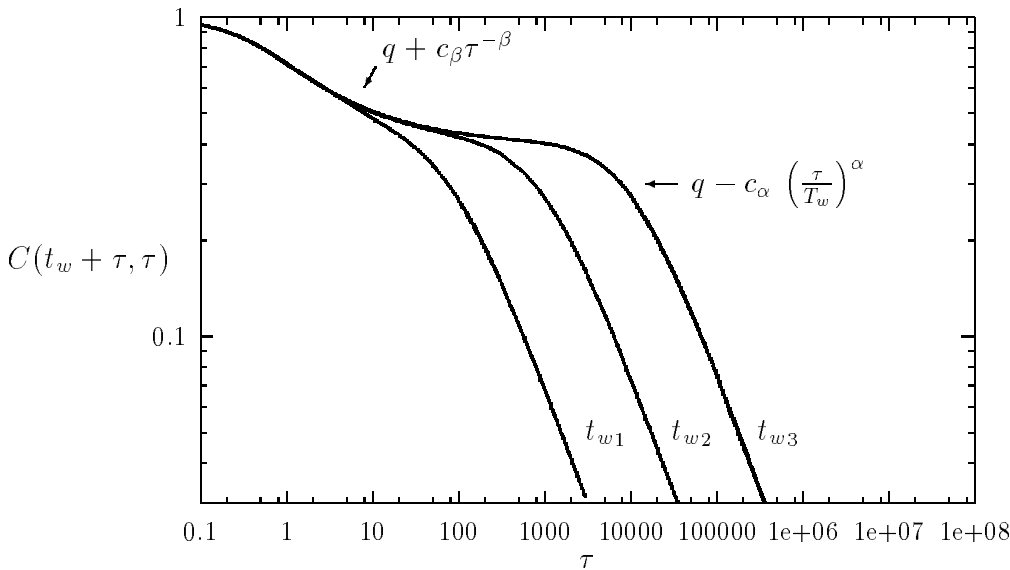

FIG. 6. Decay of the auto-correlation function below the critical temperature. $C(\tau + t_w, t_w)$ vs. $\tau$ for different waiting times, $t_{w1} < t_{w2} < t_{w3}$. $T_w = dt_w/d\ln(h(t_w))$.

## V. SUMMARY AND CONCLUSIONS

Summarizing, the major prediction of the Mode-Coupling theory of glasses for their super-cooled liquid phase is the existence of a critical temperature $T_d$ below which the correlations do not decay to zero, and above which one observes two relaxation regimes ($\alpha$ and $\beta$), characterized by a power-law behaviour with exponents related by Eq. (IV.3) – which is indeed qualitatively consistent with experimental data[3,5]. However, a quantitative comparison is difficult since experimentally, the relaxation time $\tau(T)$ *does not diverge* at $T_d$ but grows rapidly (*à la* Vogel-Fulcher) as the temperature is decreased further. In the Mode-Coupling approach, this is ascribed to some 'activated (or jump) processes' which must be taken into account in a phenomenological way. This can be rephrased differently: since we have argued that the Mode-Coupling equations are equivalent to the dynamics of a *mean-field* disordered model, it is to be expected that actual finite dimensional systems should depart from this ideal behaviour. A nucleation-like mechanism was proposed in Refs. [30], [31] to account for the smearing out of the transition in finite dimensions, but a detailed understanding of this mechanism is still lacking. This is in some sense related to the general question of assessing the quality of the MCA, and constructing perturbative schemes to move away from it[8,20].

In order to by-pass this difficulty brought about by a finite relaxation time scale below $T_d$, we propose that experiments should be done below $T_g$, so that the experimental time scales $t_w$ are much smaller than $\tau(T)$. Experimental protocols should allow one to monitor in a systematic way aging effects (i.e., the fact that the correlation function does depend on $t_w$ itself), and to obtain the curves corresponding to Fig. 6. The crucial test of Mode-Coupling theory would then be to measure both correlations (or noise) and response functions (such as dielectric properties or elastic moduli) to observe the violation of FDT and check Eqs. (IV.6) and (IV.7). It should be emphasized that most of the experimental data on supercooled liquids (and spin-glasses for that matter) can alternatively be interpreted within a phenomelogical model of 'traps'[34–36,15,37,38], where each particle diffuses in a random potential created by its neighbours. It would be interesting to understand the precise relation between this phenomenological picture and Mode-Coupling equations[39], which, as we have discussed, also describes a particle in a random potential, albeit in infinite dimension. In any case, the genuine non-trivial prediction of the Mode-Coupling theory is that the equilibration process within a 'trap' (described by the exponent $\beta$) and the aging process involving jump between traps (described, at least for small $t/t_w$, by the exponent $\alpha$) are intimately related through Eq. (IV.6). This is why we believe that its investigation is worth the experimental effort.

ACKNOWLEDGEMENTS We want to acknowledge useful discussions with C. Alba-Simionesco, A. Barrat, J.L. Barrat, R. Burioni, M. Campellone, A. Comtet, D. Dean, P. Le Doussal, G. Lozano, R. Monasson, C. Monthus and G. Parisi. L. F. C. acknowledges support from the the EU HCM grant ERB4001GT933731.

# Appendix A:

# Symmetries and the dynamical equations using the functional supersymmetric formalism

In this Appendix we review the technique of superfield notation, which is useful for three purposes:

• It provides a direct dictionary between statical and dynamical developments. There is, in this notation, a one-to-one correspondence between static and dynamical diagrams, so that one can for example talk indistinctly of 'static' and 'dynamical' MCT or SCSA.

• It makes the diagrammatic developments simpler, grouping $2^L$ ordinary diagrams of $L$ lines into a single superdiagram.

• It makes explicit a supersymmetry (SUSY) of the action (or equations of motion) which embodies the equilibrium theorems. It is then possible to see directly from the form of the action whether there is explicit or (possibly) spontaneous non-equilibrium phenomena (breaking of SUSY), and to check that an approximation scheme or an effective theory does not spoil artificially the possibility of equilibrium.

We start from a Langevin equation:

$$\frac{d\phi_\alpha}{dt} = -\frac{\partial H}{\partial \phi_\alpha} + \eta_\alpha(t) \tag{A.1}$$

where $\eta_\alpha(t)$ are Gaussian random variables with zero mean and variance $\langle \eta_\alpha(t)\, \eta_\beta(t') \rangle = 2T\, \delta_{\alpha\beta}\, \delta(t-t')$.

We now construct the Martin-Siggia-Rose[43,44] functional for the expectation value of an operator $O(\phi)$ as:

$$\langle O(\phi) \rangle = \int D[\phi]\, O(\phi) \prod_\alpha \delta\left(\frac{d\phi_\alpha}{dt} + \frac{\partial H}{\partial \phi_\alpha} - \eta_\alpha(t)\right)\, \det\left[\frac{d}{dt}\delta_{\alpha,\beta} + \frac{\partial^2 H}{\partial \phi_\alpha \partial \phi_\beta}\right] \tag{A.2}$$

where the measure $D[\phi]$ is defined as $D[\phi] \equiv \prod_\alpha D[\phi_\alpha]$.

Exponentiating the delta function through Lagrange multipliers $\hat{\phi}_\alpha(t)$ and the determinant through anticommuting variables ('ghosts') $\xi_\alpha(t), \bar{\xi}_\alpha(t)$, and averaging away the noise we obtain [‡‡]

---

[‡‡]The precise meaning of Eq. (A.4) is seen by going back to the Hilbert-space problem[42] of which it is a functional representation. Then, Eq. (A.4) represents an imaginary-time evolution operator associated with the Hamiltonian: $\sum_\alpha \hat{p}_\alpha(T\hat{p}_\alpha - i\frac{\partial H}{\partial \phi_\alpha}) + \frac{1}{2}a_\alpha^\dagger \frac{\partial^2 H}{\partial \phi_\alpha \partial \phi_\beta} a_\beta$, where $p_\alpha \equiv -i\frac{\partial}{\partial \phi_\alpha}$ The original Fokker-Planck process is recovered by restricting the problem to the zero-ghost subspace, i.e. by considering diagrams without fermionic legs. The problem of which convention (Ito or Stratonovitch) is used is simply the usual problem of factor orderings in the functional representation. Eq. (A.4) with the assumption that $\hat{\phi}_\alpha(t)\frac{\partial H}{\partial \phi_\alpha}(t)$ is understood as $\hat{\phi}_\alpha(t^+)\frac{\partial H}{\partial \phi_\alpha}(t)$ is then an unambiguous representation of the Hilbert-space problem (in the Ito convention).

$$\langle O(\phi) \rangle = \int D[\phi]\, D[\hat{\phi}]\, D[\xi]\, D[\bar{\xi}]\, O(\phi)\ \exp(-S) \tag{A.3}$$

where

$$S = \int dt \left[ \sum_\alpha (\dot{\xi}_\alpha \bar{\xi}_\alpha + \hat{\phi}_\alpha \phi_\alpha - T \hat{\phi}_\alpha^2) + \sum_\alpha \frac{\partial H}{\partial \phi_\alpha} \hat{\phi}_\alpha + \sum_{\alpha\beta} \frac{\partial^2 H}{\partial \phi_\alpha \partial \phi_\beta} \xi_\alpha \bar{\xi}_\beta \right] \tag{A.4}$$

The expression for $S$ can be written in a compact form in superspace introducing two anticommuting Grassmann variables $\theta$, $\bar{\theta}$:

$$[\theta, \bar{\theta}]_+ = \theta^2 = \bar{\theta}^2 = 0 \ . \tag{A.5}$$

The integrals over these variables are defined as:

$$\int 1 d\theta = \int 1 d\bar{\theta} = 0 \qquad \int \theta d\theta = \int \bar{\theta} d\bar{\theta} = 1 \ . \tag{A.6}$$

The fields, Lagrange multipliers and ghosts are then encoded in the (bosonic) superfield:

$$\Phi_\alpha = \phi_\alpha(t) + \bar{\theta}\, \xi_\alpha(t) + \bar{\xi}_\alpha(t)\, \theta + \hat{\phi}_\alpha(t)\, \bar{\theta}.\theta \tag{A.7}$$

Using Eqs. (A.5)-(A.6) and (A.7) one obtains, in terms of the superfields $\Phi_\alpha$

$$\langle O \rangle = \int \prod_\alpha D[\Phi_\alpha]\, O\ \exp \int d1 \left[ \frac{1}{2} \sum_\alpha \Phi_\alpha(1) D_1^{(2)} \Phi_\alpha(1) - H(\Phi(1)) \right] \tag{A.8}$$

where we have denoted $1 \equiv (\theta, \bar{\theta}, t)$, $d1 = d\theta\, d\bar{\theta}\, dt$ and the differential:

$$D^{(2)} = 2T \frac{\partial^2}{\partial\theta\, \partial\bar{\theta}} + 2\theta \frac{\partial^2}{\partial\theta\, \partial t} - \frac{\partial}{\partial t} \ . \tag{A.9}$$

The important point about expression (A.8) is that, apart for the first 'kinetic' term in the exponent and the integration over the 'time-like' coordinates $d1 = d\theta d\bar{\theta} dt$, the rest has the same form as the partition function. Furthermore, the correlation function between two superfields: $\langle \Phi_a(1)\Phi_b(2) \rangle$ (with $1 \equiv \theta_1, \bar{\theta}_1, t_1$, $2 \equiv \theta_2, \bar{\theta}_2, t_2$) encodes all correlations and response functions.

Consider now a single superfield $\Phi(1) = \phi(t) + \bar{\theta}\xi(t) + \bar{\xi}(t)\theta + \hat{\phi}(t)\bar{\theta}\theta$. For a system satisfying causality, the non-zero expectation values of the autocorrelation function are:

$$\begin{aligned} Q(1,2) &= \langle \Phi(1)\Phi(2) \rangle \\ &= \langle \phi(t_1)\phi(t_2) \rangle + (\bar{\theta}_2 - \bar{\theta}_1)\left[ \theta_2\, \langle \phi(t_1)\hat{\phi}(t_2) \rangle + \theta_1\, \langle \phi(t_2)\hat{\phi}(t_1) \rangle \right] \\ &= C(t_1,t_2) + (\bar{\theta}_2 - \bar{\theta}_1)\left[ \theta_2\, G(t_1,t_2) + \theta_1\, G(t_2,t_1) \right] \end{aligned} \tag{A.10}$$

Before going into diagrammatic computations, we need to define convolutions of two-point functions, as in:

$$Q_c(1,2) = Q_a \otimes Q_b \equiv \int d3\, Q_a(1,3) Q_b(3,2) \tag{A.11}$$

and simple products, as in:

$$Q_c(1,2) = Q_a(1,2) \bullet Q_b(1,2) \tag{A.12}$$

In what follows we shall denote with $\bullet$ any function based on usual products (e.g. $Q^{\bullet 3} \equiv Q \bullet Q \bullet Q$), and we shall omit $\otimes$ when writing convolutions (e.g. $Q^2 \equiv Q \otimes Q$).

At any step, one can go down to the 'components' of $Q$. For example, if $Q_i$ ($i = a, b$) are of the form (A.10),

$$Q_i(1,2) = C_i(t_1, t_2) + (\bar{\theta}_2 - \bar{\theta}_1)\, [\theta_2\ G_i(t_1, t_2) + \theta_1\ G_i(t_2, t_1)] \tag{A.13}$$

then $Q_c(1,2)$ obtained with the two products above is of the same form with:

$$C_c(t_1, t_2) = \int dt'\, [C_a(t_1, t')G_b(t_2, t') + G_a(t_1, t')C_b(t', t_2)]$$
$$G_c(t_1, t_2) = \int dt'\, G_a(t_1, t')G_b(t', t_2) \tag{A.14}$$

for the convolution (A.11), and:

$$C_c(t_1, t_2) = C_a(t_1, t_2) C_b(t_2, t_2)$$
$$G_c(t_1, t_2) = C_a(t_1, t_2) G_b(t_1, t_2) + G_a(t_1, t_2) C_b(t_1, t_2) \tag{A.15}$$

for the usual product (A.12).

The Mode Coupling approximation

Let us now turn to the example of equation (II.2). The Hamiltonian is given by:

$$H(\phi) = \frac{\mu(t)}{2}\phi^2 + \frac{g}{4!}\phi^4 \tag{A.16}$$

The dynamical functional then reads:

$$\langle O \rangle = \int D[\Phi]\, O(\Phi)\, \exp\left[-\int d1\, \left(\frac{1}{2}\sum_\alpha \Phi(1)\,(-D_1^{(2)} + \mu(t))\, \Phi(1) + \frac{g}{4!}\Phi^4\right)\right]. \tag{A.17}$$

Diagrams are constructed as usual. They are based on the Gaussian integrals with bare propagator

$$\mathbf{G_0} = [-D^{(2)} + \mu(t)]^{-1} \tag{A.18}$$

defined by:

$$\mathbf{G_0} \otimes [-D^{(2)} + \mu(t)] = \delta \tag{A.19}$$

where we have defined the superspace delta as

$$\delta(1-2) \equiv \delta(t_1 - t_2)(\bar{\theta}_1 - \bar{\theta}_2)(\theta_1 - \theta_2)\,. \tag{A.20}$$

With these definitions, a 'superdiagram' is obtained just as an ordinary diagram, with now the labels $1,2$ encoding both the times and the Grassmann coordinates, indicating the 'components' $C, G$ of $Q$. Each line of the superdiagram stands for a line of field-field $\langle \phi(t_1)\phi(t_2) \rangle$ contraction ($\sim C$) and a line of field-noise $\langle \phi(t_1)\eta(t_2) \rangle$ contraction ($\sim G$). The Dyson equations (II.7)-(II.6) are both encoded in (see Fig. 7):

$$Q = \mathbf{G_0} + \mathbf{G_0} \otimes \mathbf{\Sigma} \otimes Q \tag{A.21}$$

The (super) mass-operator $\mathbf{\Sigma}$ can be now calculated within the mode-coupling approximation, which again consists in neglecting vertex corrections. It is given by:

$$\mathbf{\Sigma}(1,2) = \frac{g^2}{6} Q^{\bullet 3}(1,2) \tag{A.22}$$

Introducing this into (A.21), and multiplying (in the sense of convolution) by $\mathbf{G_0}^{-1}$, we obtain

$$-D_1^{(2)} Q(1,2) = -\mu(t) Q(1,2) + \delta(1-2) + \frac{g^2}{6} \int d3 \, [Q(1,3)]^{\bullet 3} Q(3,2) \tag{A.23}$$

This is the equation of motion. For a general non-linear coupling, one similarly obtains

$$-D_1^{(2)} Q(1,2) = -\mu(t) Q(1,2) + \delta(1-2) + \int d3 \, \hat{\mathcal{V}}'^{\bullet}(Q(1,3)) \, Q(3,2) \tag{A.24}$$

which, in components, is nothing but Eqs. (II.27) and (II.28).

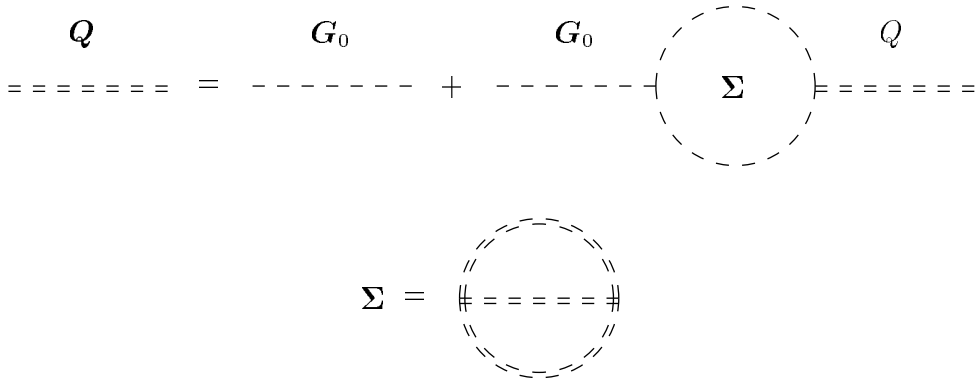

**Figure 7**. Diagrammatic representation in SUSY formalism of the Dyson Equations. In this notation one encodes the $C - G$ diagrams of Fig. 3 in only one super-diagram.

## The disordered model for the MCA

One can reconstruct an action functional of which (A.24) is a stationary point. To do this, we multiply (A.24) to the right by $Q^{-1}$:

$$(D_1^{(2)} - \mu(t))\delta(1-2) + Q^{-1}(1,2) + \hat{\mathcal{V}}'^{\bullet}(Q)(1,2) = 0 \qquad (A.25)$$

Which can be written as:

$$\frac{\delta S}{\delta Q} = 0$$

$$2S = \int d1\, d2\, \left[(-D_1^{(2)} + \mu(t))\, Q(1,2) - \hat{\mathcal{V}}^{\bullet}(Q)\right] - \text{Tr Ln}[Q] \qquad (A.26)$$

Let us now see how these are the exact equations of motion for a disordered system. The dynamical generating functional reads in superspace notation

$$\int D[\Phi]\, \exp\left[-\int d1\, \left(\sum_{\alpha}^{N} \frac{1}{2} \Phi_\alpha(1)\, (-D_1^{(2)} + \mu(t))\, \Phi_\alpha(1) + H_J[\Phi]\right)\right], \qquad (A.27)$$

with the disordered Hamiltonian

$$H_J[\Phi] = g \sum_{r \geq 2}^{\infty} F_r \sum_{\alpha_1 < \ldots < \alpha_{r+1}} J_{\alpha_1 \ldots \alpha_{r+1}}\, \Phi_{\alpha_1} \ldots \Phi_{\alpha_{r+1}} \qquad (A.28)$$

correlated as in (II.25).

Averaging over the couplings (see Eq.(II.25)), we obtain

$$\int D[\Phi]\, \exp\left[-\int d1\, (\frac{1}{2}) \sum_\alpha \Phi_a(1)\, (-D_1^{(2)} + \mu(t))\, \Phi_\alpha(1)\right]$$
$$\exp\left[\frac{N}{2} \int d1 d2\, \hat{\mathcal{V}}^{\bullet}\left(\frac{\Phi(1) \cdot \Phi(2)}{N}\right)\right]. \qquad (A.29)$$

$$(A.30)$$

Introducing the order parameter

$$Q(1,2) = \frac{1}{N} \sum_\alpha \Phi_\alpha(1)\Phi_\alpha(2) \qquad (A.31)$$

through

$$1 = \int D[Q]\, D[\hat{Q}] \exp\left[\frac{1}{2} \int d1\, d2\, \left(NQ(1,2)\hat{Q}(1,2) - \hat{Q}(1,2) \sum_\alpha \Phi_\alpha(1)\Phi_\alpha(2)\right)\right] \qquad (A.32)$$

yields

$$\int D[\Phi]D[Q]D[\hat{Q}]\, \exp\left[-\frac{1}{2} \int d1 d2\, \left(NQ(1,2)\hat{Q}(1,2) - \hat{\mathcal{V}}^{\bullet}(Q(1,2))\right.\right.$$
$$\left.\left. - \sum_\alpha \Phi_\alpha(1)\, (-D_1^{(2)} + \mu(t))\, \delta(1-2) + \hat{Q}(1,2))\, \Phi_\alpha(2)\right)\right]. \qquad (A.33)$$

We can now make the shift:

$$\overline{Q} = (-D_1^{(2)} + \mu(t))\,\delta(1-2) + \hat{Q}(1,2) \tag{A.34}$$

and the integration over $\Phi$, to obtain

$$\int D[Q]D[\overline{Q}] \exp\left[-\frac{N}{2}\int d1 d2\,\left(Q(1,2)\overline{Q}(1,2) + (-D_1^{(2)} + \mu(t))\,Q(1,2) - \hat{\mathcal{V}}^\bullet(Q(1,2))\right)\right]$$
$$\exp\left[-\frac{N}{2}\,\mathrm{Tr}\,\mathrm{Ln}\,\overline{Q}\right]. \tag{A.35}$$

Using saddle-point evaluation, we can eliminate $\overline{Q}$, and obtain:

$$\int D[Q]\,\exp(-NS)$$
$$2S = \int d1\,d2\,\left[\,(-D_1^{(2)} + \mu(t))\,Q(1,2) - \hat{\mathcal{V}}^\bullet(Q(1,2))\,\right] - \mathrm{Tr}\,\mathrm{Ln}[Q] \tag{A.36}$$

The saddle-point equation over $Q$ yields (A.25).

<u>SCSA and the Bernasconi model</u>

In a similar way we show that the equations of motion for the disordered Bernasconi model (III.9) coincide with the equations arising from the SCSA applied to the $\phi^4$ model. The generating function for the dynamics reads, in superspace notation:

$$\int D[\Phi]\exp\left[-\int d1\,\left(\frac{1}{2}\sum_{\alpha=1}^N \Phi_\alpha(1)\,(-D_1^{(2)} + \mu(t))\,\Phi_\alpha(1) + \frac{J_o}{2N}\sum_{\lambda=1}^N\left(\sum_{\alpha<\beta=1}^N J_{\alpha,\beta}^\lambda \Phi_\alpha \Phi_\beta\right)^2\right)\right]. \tag{A.37}$$

Making a Gaussian transformation by means of *superfields* $\sigma_\lambda$:

$$\int D[\Phi]\,D[\sigma]\,\exp\left[-\int d1\,\left(\frac{1}{2}\sum_{\lambda=1}^N \Phi_\alpha(1)\,(-D_1^{(2)} + \mu(t))\,\Phi_\alpha(1)\right.\right.$$
$$\left.\left.+ i\sqrt{\frac{J_o}{N}}\sum_{\alpha<\beta,\lambda} J_{\alpha,\beta}^\lambda \Phi_\alpha \Phi_\beta \sigma_\lambda + \frac{1}{2}\sum_\lambda \sigma_\lambda^2\right)\right]. \tag{A.38}$$

Averaging over the couplings:

$$\int D[\Phi]D[\sigma]\,\exp\left[-\int d1\,\left(\frac{1}{2}\sum_{\alpha=1}^N \Phi_\alpha(1)\,(-D_1^{(2)} + \mu(t))\,\Phi_\alpha(1)\right)\right]$$
$$\exp\left[-\frac{1}{2}\int d1 d2\,\left(\left(2J_o\left(\frac{1}{N}\sum_\alpha \Phi_\alpha(1)\Phi_\alpha(2)\right)^2 + \delta(1-2)\right)\sum_\lambda \sigma_\lambda(1)\sigma_\lambda(2)\right)\right]. \tag{A.39}$$

Note that putting $\overline{J_{\alpha,\beta}^\lambda} = O(1/N)$ creates an infinitely strong antiferromagnetic force. One can then neglect in the previous expression a term $\int d1d2 \left(\sum_\alpha \Phi_\alpha(1)/\sqrt{N}\right)^2 \left(\sum_\beta \Phi_\beta(2)/\sqrt{N}\right)^2$ since it is of order $O(1)$ and hence much smaller than the other $O(N)$ terms. One can just as well put $\overline{J^\lambda_{\alpha,\beta}} = 0$ and see directly how this term disappears.

Introducing two order parameters $Q_\Phi$ and $Q_\sigma$ as in (A.31)-(A.32)

$$N\, Q_\Phi(1,2) \equiv \sum_\alpha \Phi_\alpha(1)\Phi_\alpha(2) \qquad N\, Q_\sigma(1,2) \equiv \sum_\lambda \sigma_\lambda(1)\sigma_\lambda(2) , \qquad (A.40)$$

we obtain

$$\int D[\Phi]D[\sigma]D[Q_\Phi]D[\hat{Q}_\Phi]D[Q_\sigma]D[\hat{Q}_\sigma]$$
$$\exp\left[-\int d1d2 \left(\frac{N}{2} Q_\Phi(1,2)\hat{Q}_\Phi(1,2) + \frac{1}{2}\left(NQ_\sigma(1,2) - \sum_\lambda \sigma_\lambda(1)\sigma_\lambda(2)\right)\hat{Q}_\sigma(1,2)\right.\right.$$
$$\left.\left.-\frac{1}{2}\sum_{\alpha=1}^N \Phi_\alpha(1)\left(\left[-D_1^{(2)} + \mu(t)\right]\delta(1-2) + \hat{Q}_\Phi(1,2)\right)\Phi_\alpha(2)\right)\right]$$
$$\exp\left[-\frac{1}{2}\int d1d2\, \left(\left(2J_o\, Q_\Phi^{\bullet 2}(1,2) + \delta(1-2)\right) Q_\sigma(1,2)\right)\right] \qquad . (A.41)$$

Making a shift (A.34) over $Q_\Phi$ and the integration over $\Phi$ and $\sigma$:

$$\int D[Q_\Phi]D[\overline{Q}_\Phi]D[Q_\sigma]D[\overline{Q}_\sigma]$$
$$\exp\left[-\frac{N}{2}\int d1d2\, \left(Q_\Phi(1,2)\overline{Q}_\Phi(1,2) + Q_\sigma(1,2)\overline{Q}_\sigma(1,2) + (-D_1^{(2)} + \mu(t))\, Q_\Phi(1,2)\right)\right]$$
$$\exp\left[-\frac{N}{2}\left(\mathrm{TrLn}\overline{Q}_\Phi + \mathrm{TrLn}\hat{Q}_\sigma\right) - \frac{N}{2}\int d1d2\, \left(\left(2J_o\, Q_\Phi^{\bullet 2}(1,2) + \delta(1-2)\right) Q_\sigma(1,2)\right)\right] . (A.42)$$

Eliminating $\overline{Q}_\Phi$ and $\hat{Q}_\sigma$ through saddle point evaluation, we obtain the action:

$$\int D[Q_\Phi]D[Q_\sigma]\, \exp(-NS)$$
$$2S = \int d1d2\, \left[(-D_1^{(2)} + \mu(t))Q_\Phi(1,2) + [2J_o\, Q_\Phi^{\bullet 2} + \delta] \bullet Q_\sigma\right] - \mathrm{TrLn}[Q_\Phi] - \mathrm{TrLn}[Q_\sigma] . \quad (A.43)$$

The equations of motion are obtained by saddle point over $Q_\Phi$ and $Q_\sigma$

$$0 = (-D_1^{(2)} + \mu(t))\, \delta(1-2) + 4J_o\, Q_\Phi(1,2)Q_\sigma(1,2) - Q_\Phi^{-1}(1,2) \qquad (A.44)$$
$$Q_\sigma^{-1}(1,2) = (2J_o\, Q_\Phi^{\bullet 2} + \delta)(1,2) \qquad (A.45)$$

and multiplying the first equation by $Q_\Phi$ and the second by $Q_\sigma$ we finally get:

$$0 = (-D_1^{(2)} + \mu(t))\, Q_\Phi(1,2) + 4J_o\, ((Q_\Phi \bullet Q_\sigma)Q_\Phi)(1,2) - \delta(1-2) \qquad (A.46)$$
$$\delta(1-2) = ((2J_o\, Q_\Phi^{\bullet 2} + \delta)Q_\sigma)(1,2) . \qquad (A.47)$$

These equations when written in components become equivalent to the equations of motion for the SCSA, Eqs.(III.4)-(III.8).

The generalization of what we have done to several modes that derive from an energy is straightforward.

We can now see the formal difference between MCA and SCSA. Both the MCA and the SCSA equation for $Q_\Phi$ are of the form:

$$(-D_1^{(2)} + \mu(t))Q(1,2) = \delta(1-2) + \int d3\, m[Q](1,3)Q(3,2) \tag{A.48}$$

The kernel $m[Q]$ is a *function* $m[Q](1,3) = \hat{\mathcal{V}}'^\bullet(Q(1,3))$ for MCA, while it is a non-local *functional* $m[Q_\Phi] = 4J_o(2J_o Q_\Phi^{\bullet 2} + \delta)^{-1} \bullet Q_\Phi$ in SCSA.

Symmetries

Let us finally turn to the question of the symmetries associated with the equilibrium theorems. The SUSY group is generated by three operators[40–42]:

$$\bar{D}' = T\frac{\partial}{\partial \theta} + \bar{\theta}\frac{\partial}{\partial t}$$
$$D' = \frac{\partial}{\partial \bar{\theta}}$$
$$[D', \bar{D}']_+ = \frac{\partial}{\partial t} \tag{A.49}$$

$$D'^2 = \bar{D}'^2 = 0 \tag{A.50}$$

We can construct a version of this group that acts on two-point (in general $n$-point) functions, as:

$$\mathbf{D}' = D'(1) + D'(2)$$
$$\bar{\mathbf{D}}' = \bar{D}'(1) + \bar{D}'(2)$$
$$[\mathbf{D}', \bar{\mathbf{D}}']_+ = \frac{\partial}{\partial t_1} + \frac{\partial}{\partial t_2} \tag{A.51}$$

The meaning of the three generators can be understood when they are made to act on a correlation function. Firstly, causality plus probability conservation imply (irrespective of equilibration):

$$\mathbf{D}'Q(1,2) = 0 \tag{A.52}$$

(This was already assumed in selecting the non-zero terms in (A.10)). The other two generators:

$$\left(\frac{\partial}{\partial t_1} + \frac{\partial}{\partial t_2}\right)Q(1,2) = 0 \quad \rightarrow \quad TTI\,,$$
$$\bar{\mathbf{D}}'Q(1,2) = 0 \quad \rightarrow \quad FDT\,.$$

The question of non-equilibrium can now be stated as follows:
• Systems with *explicit* non-equilibrium have a dynamical action (or dynamical equations of motion) that break SUSY explicitly.

If the system is a priori able to achieve equilibrium, then SUSY is not explicitly broken. Then, two things may happen:

• The effect of the initial conditions is a *finite* transient $\sim t_{eq}$ in which TTI and FDT do not hold. In this language, SUSY is unbroken by the boundary conditions.

• If the system never achieves equilibrium, as in the case of the low-temperature version of the MCT equations $t_{eq} \to \infty$, the effect of the initial conditions is then to break SUSY[42,45] (violate TTI and FDT) well within the 'bulk' of times. SUSY is then spontaneously broken.

The initial conditions play for SUSY (FDT and TTI) the same role played in ordinary symmetry-breaking by space boundary conditions: if the symmetry is spontaneously broken their effect extends away from them.

Hence, if one is treating a system like a spin or structural glass within an approximation (or a phenomenological model), one must make sure that the resulting theory does not break SUSY *explicitly*, otherwise one may be introducing non-equilibrium by hand.

## Appendix B:

## Derivation of the SCSA equations from a disordered Bernasconi model

In this Appendix, we derive the dynamical equations corresponding to the disordered Bernasconi model using standard functional methods. For a more compact derivation using supersymmetric functional methods, see Appendix A. Following Ref. [11], we will define the disordered version of the Bernasconi model by the following Hamiltonian:

$$\mathcal{H} = \frac{J_0}{2N} \sum_{\lambda}^{N} [\sum_{\alpha<\beta}^{N} J_{\alpha,\beta}^{\lambda} \phi_\alpha \phi_\beta]^2 + \frac{\mu}{2} \sum_{\alpha} \phi_\alpha^2 + \frac{g'}{4N}[\sum_{\alpha} \phi_\alpha^2]^2 \qquad (B.1)$$

For each $\lambda$ independently, $J_{\alpha,\beta}^{\lambda} = J_{\beta,\alpha}^{\lambda}$ is equal to 1 with probability $\frac{1}{N}$ and zero otherwise. Each $\lambda$ thus corresponds to a certain 'pairing' of the 'sites' $\{\alpha\}$. For $J_0 > 0$, the ground state corresponds to a configuration of the $\phi_\alpha$ which simultaneously minimizes all the 'partial correlation' $S_\lambda$ defined for each pairing as

$$S_\lambda(t) \equiv \frac{1}{\sqrt{N}} \sum_{\alpha<\beta}^{N} J_{\alpha,\beta}^{\lambda} \phi_\alpha \phi_\beta \qquad (B.2)$$

The problem is extremely frustrated if the $\phi_\alpha$ are Ising spins, and becomes trivial if the $\phi_\alpha$ are unconstrained continuous variables. The model defined in (B.1) is intermediate, since, as usual, a certain amount of constraint is enforced by the terms proportional to $\mu$ and $g'$.

The dynamical equations read:

$$\frac{\partial \phi_\alpha}{\partial t} = -\frac{\partial \mathcal{H}}{\partial \phi_\alpha} + \eta_\alpha(t) \qquad \langle \eta_\alpha(t) \eta_\beta(t') \rangle = 2T \delta_{\alpha,\beta} \delta(t-t') \qquad (B.3)$$

Now, the two identities (B.2),(B.3) can be written in a convenient functional way by expressing the $\delta$ functions in Fourier space. (We set $g' = 0$ for simplicity, and will discuss the modifications induced by $g' \neq 0$ at the end of the calculation). We thus write:

$$\int \prod_t [\prod_\alpha d\phi_\alpha(t) d\hat\phi_\alpha(t)][\prod_\lambda dS_\lambda(t) d\hat S_\lambda(t)] \exp -\int dt [\sum_\lambda \hat S_\lambda(t) \{S_\lambda(t) - \frac{1}{\sqrt{N}} \sum_{\alpha<\beta} J_{\alpha,\beta}^\lambda \phi_\alpha(t) \phi_\beta(t)\} +$$

$$+ \sum_\alpha \hat\phi_\alpha(t) \{\dot\phi_\alpha(t) + \mu \phi_\alpha(t) + \frac{J_0}{\sqrt{N}} \sum_{\lambda,\beta} S_\lambda(t) J_{\alpha,\beta}^\lambda \phi_\beta(t) + \eta_\alpha(t)\}] \equiv 1 \qquad (B.4)$$

(The Jacobian associated with the change of variables is equal to 1 if one uses the Ito prescription). Averaging over the (Gaussian) thermal noise amounts to replace $\hat\phi_\alpha(t)\eta_a(t)$ by $T\hat\phi_\alpha(t)^2$. The average over the $J_{\alpha,\beta}^\lambda$ can also be performed and leads, for $N$ large, to the following expression:

$$\exp \sum_\lambda \sum_{\alpha<\beta} [-\frac{1}{N^{3/2}} \int dt (-\hat S_\lambda \phi_\alpha \phi_\beta + J_0 S_\lambda \hat\phi_\alpha \phi_\beta + J_0 S_\lambda \phi_\alpha \hat\phi_\beta)(t)+$$

$$\frac{1}{2N^2}\int dt dt'(-\hat{S}_\lambda \phi_\alpha \phi_\beta + J_0 S_\lambda \hat{\phi}_\alpha \phi_\beta + J_0 S_\lambda \phi_\alpha \hat{\phi}_\beta)(t)(-\hat{S}_\lambda \phi_\alpha \phi_\beta + J_0 S_\lambda \hat{\phi}_\alpha \phi_\beta + J_0 S_\lambda \phi_\alpha \hat{\phi}_\beta)(t')]$$
(B.5)

The second is a sum of $N^3$ terms with a positive mean, and is thus of order $N$, while the first term is a sum of $N^3$ terms of random sign, and is of order 1, which we neglect.

The next step is to define six 'correlation functions', associated to the fields $\phi_\alpha, \hat{\phi}_\alpha, S_\lambda, \hat{S}_\lambda$. Let us introduce:

$$C_\phi(t,t') = N^{-1} \sum_\alpha \phi_\alpha(t)\phi_\alpha(t') \quad G_\phi(t,t') = N^{-1} \sum_\alpha \phi_\alpha(t)\hat{\phi}_\alpha(t')$$

$$Z_\phi(t,t') = N^{-1} \sum_\alpha \hat{\phi}_\alpha(t)\hat{\phi}_\alpha(t') \tag{B.6}$$

and similarly for the $C_S, G_s, Z_S$. The expectation values of $G$ are actually response functions[44] and those of $Z$ are in fact zero[44] but $Z$ must be kept in the intermediate steps of the calculation. Again, these identities are expressed as $\delta$ functions, introducing six new 'conjugate' variables $N\hat{C}_{\phi,S}, N\hat{G}_{\phi,S}, N\hat{Z}_{\phi,S}$. The 'interaction term' (B.5), expressed in terms of the $C,G,Z$ simply reads:

$$\exp \frac{N}{2}\int_{t'\leq t} dtdt'[Z_S(t,t')C_\phi(t,t')^2 + 2J_0^2 C_S(t,t')Z_\phi(t,t')C_\phi(t,t')-$$

$$2J_0 G_S(t',t)G_\phi(t,t')C_\phi(t,t') - 2J_0 G_S(t,t')G_\phi(t',t)C_\phi(t,t')+$$

$$+2J_0^2 G_\phi(t',t)G_\phi(t,t')C_S(t,t')] \tag{B.7}$$

The point now is that all the terms containing $C,G,Z$ are proportional to $N$, and can be treated within a saddle point approximation which becomes exact when $N$ is large. The saddle point equations read:

$$\frac{\partial}{\partial C_\phi} = 0 \longrightarrow \hat{C}_\phi = Z_S C_\phi + J_0^2 Z_\phi C_S - J_0 G_S^\dagger G_\phi - J_0 G_S G_\phi^\dagger \tag{B.8-a}$$

$$\frac{\partial}{\partial G_\phi} = 0 \longrightarrow \hat{G}_\phi = -J_0 G_S^\dagger C_\phi + J_0^2 G_\phi^\dagger C_S \tag{B.8-b}$$

$$\frac{\partial}{\partial Z_\phi} = 0 \longrightarrow \hat{Z}_\phi = J_0^2 C_S C_\phi \tag{B.8-c}$$

$$\frac{\partial}{\partial C_S} = 0 \longrightarrow \hat{C}_S = J_0^2 G_\phi^\dagger G_\phi - J_0^2 Z_\phi C_\phi \tag{B.8-d}$$

$$\frac{\partial}{\partial G_S} = 0 \longrightarrow \hat{G}_S = -J_0 G_\phi^\dagger C_\phi \tag{B.8-e}$$

$$\frac{\partial}{\partial Z_S} = 0 \longrightarrow \hat{Z}_S = \frac{C_\phi^2}{2} \tag{B.8-e}$$

where we have dropped the arguments $(t,t')$ when they appear in the correct order $(t>t')$, and indicated with a † when they appear in reversed order.

From their physical interpretation (see, e.g. Ref. [44]), one expects that $Z_S = Z_\phi \equiv 0$ and $G_\phi = G_S = 0$ for $t < t'$. The saddle point estimate of Eq. (B.4) averaged over the $J$'s then finally reads:

$$\int \prod_t [\prod_\alpha d\phi_\alpha(t) d\hat{\phi}_\alpha(t)][\prod_\lambda dS_\lambda(t) d\hat{S}_\lambda(t)] \exp -\int dt [\sum_\lambda \hat{S}_\lambda(t)\{S_\lambda(t)+$$

$$\int_{t' \leq t} dt' J_0 C_\phi G_\phi S_\lambda(t') - \frac{1}{2} C_\phi^2 \hat{S}_\lambda(t')\} + \sum_\alpha \hat{\phi}_\alpha(t)\{\dot{\phi}_\alpha(t) + \mu \phi_\alpha(t)+$$

$$+ \int_{t' \leq t} dt' J_0 (C_\phi G_S - J_0 C_S G_\phi) \phi_\alpha(t') - (T\delta(t-t') + J_0^2 C_\phi C_s) \hat{\phi}_\alpha(t')\}] \equiv 1 \quad (B.9)$$

This last equation is easy to interpret by comparison with Eq. (B.4) – it is simply the Fourier representation of some $\delta$ functions implementing the following equations of motion:

$$G_{0\phi}^{-1} \phi(t) = J_0 \int_0^t dt' (J_0 C_S G_\phi - C_\phi G_S) \phi(t') + \xi(t) \quad (B.10)$$

with $G_{0\phi}^{-1} = \frac{\partial}{\partial t} + \mu$ and $\langle \xi(t)\xi(t') \rangle = 2T\delta(t-t') + J_0^2 C_S C_\phi$ and

$$G_{0S}^{-1} S(t) = -J_0 \int_0^t dt' C_\phi G_\phi S(t') + \zeta(t) \quad (B.11)$$

with $G_{0S} = \delta(t - t')$ and $\langle \zeta(t)\zeta(t') \rangle = \frac{1}{2} C_\phi^2$. Note that $C, G$ are self-consistently determined. Hence, comparing with Eqs.(III.10,III.11), we indeed see that the SCSA for the usual $\phi^4$ theory are the exact equations describing the random Bernasconi model defined by Eq. (B.1), provided one identifies $J_0$ with the $\phi^4$ coupling, $\frac{2g}{3!}$. The only missing part is the 'tadpole' contribution, which can be easily added by choosing a suitable value of $g'$ in Eq. (B.1), since this last term only adds in the equation of motion of $\phi_\alpha$ a non fluctuating contribution $-g'\phi_a C_\phi(t,t)$ (for $N$ large).

One should note that, as emphasized in the text, $g > 0$ in the original $\phi^4$ theory corresponds to $J_0 > 0$, and hence to a well defined (bounded from below) Hamiltonian $\mathcal{H}$. The dynamical SCSA equations are thus expected to have sensible solutions for all values of parameters, contrarily to the direct MCA.

# REFERENCES

[1] For a review, see e.g. W. D. McComb, *The Physics of Fluid Turbulence*, Oxford Science Publications, 1990. R. H. Kraichnan and S. Chen; Physica D **37** (1989) 160.

[2] T. Halpin-Healey and Y.C. Zhang; Phys. Rep. **254** (1995) 217.

[3] For reviews, see W. Götze, in *Liquids, freezing and glass transition*, Les Houches 1989, JP Hansen, D. Levesque, J. Zinn-Justin Editors, North Holland. see also W. Götze, L. Sjögren, Rep. Prog. Phys. **55** (1992) 241.

[4] There are so many papers using this technique that we only give a very subjective selection of them: R. Kraichnan; J. Fluid. Mech. **5** 497 (1959). K. Kawasaki; Ann. Phys. **61** (1970) 1. H. Van Beijeren, R. Kutner and H. Spohn; Phys. Rev. Lett **54** (1985) 2026. In the context of the KPZ equation, see: J. Krug; Phys. Rev. **A36** (1987) 5465 J.P. Bouchaud and M.E. Cates; Phys. Rev. **E47** (1993) R1455; Erratum **E48** (1993) 635. M.A. Moore, T. Blum, J. Doherty, M. Marsili, J.P. Bouchaud and Ph. Claudin; Phys. Rev. Lett. Phys. Rev. Lett **74** (1995) 4257. For related ideas, see also: M. Schwartz and S.F. Edwards; Europhys. Lett **20** (1992) 301. T. Blum and A.J. McKane; cond-mat **9505038**.

[5] For an enlightening introduction to the experimental controversy, see the series of Comments in Phys. Rev. **E**: X.C. Zeng, D. Kivelson and G. Tarjus; Phys. Rev. **E50** (1994) 1711. P. K. Dixon, N. Menon and S. R. Nagel; Phys. Rev. **E50** (1994) 1717. H. Z. Cummins and G. Li; Phys. Rev. **E50** (1994) 1720 and references therein, in particular, H. Z. Cummins, W.M. Du, M. Fuchs, W. Götze, S. Hildebrand, A. Latz, G. Li and N.J. Tao, Phys. Rev. **E47** (1993) 4223.

[6] R. Kraichnan; J. Fluid. Mech. **7** (1961) 124.

[7] A. Bray; Phys. Rev. Lett. **32** (1974) 1413. For recent developments in the context of disordered systems, see M. Mézard and A. P. Young; Europhys. Lett. **18** (1992) 653. M. Mézard and R. Monasson; Phys.Rev. **B50** (1994) 7199.

[8] C.Y. Mou and P. B. Weichman; Phys. Rev. Lett. **70**, 1101, (1993).

[9] S. Franz and J. Hertz; Phys. Rev. Lett. **74** (1995) 2114.

[10] T. R. Kirkpatrick and D. Thirumalai; J. Phys. **A22** (1989) L149.

[11] J.P. Bouchaud and M. Mézard; J. Phys. I (France) **4** (1994) 1109. E. Marinari, G. Parisi and F. Ritort; J. Phys. **A27** (1994) 7615; J. Phys. **A27** (1994) 7647.

[12] L. F. Cugliandolo, J. Kurchan, G. Parisi and F.Ritort; Phys. Rev. Lett. **74** (1995) 1012. P. Chandra, L. Ioffe and D. Sherrington; cond-mat **9502018**. P. Chandra, M. Feigelmann and L. Ioffe; preprint cond-mat **9509022**.

[13] E. Vincent, J. Hammann and M. Ocio; p. 207 in *Recent Progress in Random Magnets*, D.H. Ryan Editor, (World Scientific Pub. Co. Pte. Ltd, Singapore 1992).

[14] L. C. E. Struik; *Physical aging in amorphous polymers and other materials* (Elsevier, Houston, 1978).

[15] J. P. Bouchaud; J. Phys. I (France) **2** (1992) 1705. J. P. Bouchaud and D. S. Dean; J. Phys. I (France) **5** (1995) 265.

[16] L. F. Cugliandolo and J. Kurchan; Phys. Rev. Lett. **71** (1993) 173; J. Phys. **A27** (1994) 5749.

[17] S. Franz and M. Mézard; Europhys. Lett. **26** (1994) 209; Physica **A209** (1994) 1.

[18] L. F. Cugliandolo and P. Le Doussal, *Large time off-equilibrium dynamics of a particle diffusing in a random potential*, cond-mat **9505112**, to be published in Phys. Rev. **E**.

L. F. Cugliandolo, J. Kurchan and P. Le Doussal, preprint cond-mat **9509009**.

[19] M. Mézard, G. Parisi and M.A. Virasoro, *Spin glass theory and beyond*, chapter VI, (World Scientific, Singapore) 1987.

[20] G. Eynick, Phys. Rev. **E49** (1994) 3990.

[21] J. Bernasconi, J. Physique **48** (1987) 559.

[22] D.J. Amit and D.V. Roginsky; J. Phys. **A12** (1979) 689.

[23] J.P. Doherty, A.J. Bray, J.M. Kim and M.A. Moore, Phys. Rev. Lett. **72** (1994) 2041.

[24] For a review, see: A. J. Bray, Adv. in Phys. **43** (1994) 357

[25] M. Campellone et al.; in preparation.

[26] S. Gauthier, M.E. Brachet, J.N. Fournier, J. Phys. A **14** (1981) 2969

[27] H.J. Sommers and K.H. Fischer; Z. Phys. **B58**, 125 (1985).

[28] T. R. Kirkpatrick and D. Thirumalai; Phys. Rev. **B36** (1987) 5388.

[29] H. Kinzelbach and H. Horner; J. Phys. I (France) **3** (1993) 1329; J. Phys. I (France) **3** (1993) 1901.

[30] T. R. Kirkpatrick, D. Thirumalai and P. G. Wolynes; Phys. Rev. **A40** (1989) 1045 and references therein.

[31] G. Parisi; *Slow dynamics in glasses*, cond-mat **941115** and **9412034**.

[32] M. Mézard, G. Parisi, J. Physique I **1** 809 (1991); J.Phys. **A23** L1229 (1990)

[33] For a review, see the interesting series of papers in Science, **267** (1995) 1924.

[34] H. Bassler; Phys. Rev. Lett. **58** (1987) 767, V. Arkhipov, H. Bassler, Phys. Rev. E **52** (1995) 1227 and references therein.

[35] T. Odagaki, J. Matsui and Y. Hiwatari; Physica **A204** 464 (1994).

[36] S. Gomi and F. Yonezawa; **74** (1995) 4125.

[37] A. Barrat and M. Mézard; J. Phys. I (France) **5** (1995) 941.

[38] J.P. Bouchaud, A. Comtet and C. Monthus; *On a dynamical model of glasses*, cond-mat **9506027**, to appear in Journal de Physique I (Dec. 1995) and C. Monthus, J.P. Bouchaud, in preparation.

[39] Such a program was attempted, in a rather formal way, in L. Sjögren, Z. Phys. B **74** (1989) 353.

[40] E. Gozzi; Phys.Rev. **D30** (1984) 1218.

[41] J.Zinn Justin; *"Quantum Field Theory and Critical Phenomena"* Clarendon Press, Oxford (1989).

[42] J. Kurchan; J.Phys. I (France) **I** (1992) 1333.

[43] P.C. Martin, E.D. Siggia and H.A. Rose; Phys. Rev. **A8** (1978) 423.

[44] C. de Dominicis and L. Peliti; Phys. Rev. **B18** (1978) 353.

[45] L.F.Cugliandolo and J. Kurchan, Phil. Magaz. **B71** (1995) 50.